\documentclass[5p,twocolumn,times,number]{elsarticle}

\usepackage{textcomp}
\usepackage{graphicx}
\usepackage{amsmath}
\usepackage[colorlinks,citecolor=blue,urlcolor=blue,linkcolor=blue]{hyperref}

\begin{document}

\begin{frontmatter}

\title{Charged particle timing at sub-25 picosecond precision: the PICOSEC detection concept}

\author[CEA]{F.J.~Iguaz\corref{cor}}
\ead{iguaz@cea.fr}
\author[CERN]{J.~Bortfeldt}
\author[CERN]{F.M.~Brunbauer}
\author[CERN]{C.~David}
\author[CEA]{D.~Desforge}
\author[NSCR]{G.~Fanourakis}
\author[CERN]{J.~Franchi}
\author[LIPP]{M.~Gallinaro}
\author[HIP]{F.~Garc\'ia}
\author[CEA]{I.~Giomataris}
\author[IGFAE]{D.~Gonz\'alez-D\'iaz}
\author[LIDYL]{T.~Gustavsson}
\author[CEA]{C.~Guyot}
\author[CEA]{M.~Kebbiri}
\author[CEA]{P.~Legou}
\author[USTC]{J.~Liu}
\author[CERN]{M.~Lupberger}
\author[CEA]{O.~Maillard}
\author[AUTH]{I.~Manthos}
\author[CERN]{H.~M\"uller}
\author[AUTH]{V.~Niaouris}
\author[CERN]{E.~Oliveri}
\author[CEA]{T.~Papaevangelou}
\author[AUTH]{K.~Paraschou}
\author[LIST]{M.~Pomorski}
\author[USTC]{B.~Qi}
\author[CERN]{F.~Resnati}
\author[CERN]{L.~Ropelewski}
\author[AUTH]{D.~Sampsonidis}
\author[CERN]{T.~Schneider}
\author[CEA]{P.~Schwemling}
\author[CEA]{L.~Sohl}
\author[CERN]{M.~van~Stenis}
\author[CERN]{P.~Thuiner}
\author[NTUA]{Y.~Tsipolitis}
\author[AUTH]{S.E.~Tzamarias}
\author[RD51]{R.~Veenhof\fnref{Veenhof}}
\author[USTC]{X.~Wang}
\author[CERN]{S.~White\fnref{Virginia}}
\author[USTC]{Z.~Zhang}
\author[USTC]{Y.~Zhou}

\cortext[cor]{Corresponding author}

\address[CEA]{IRFU, CEA, Universit\'e Paris-Saclay, F-91191 Gif-sur-Yvette, France}
\address[CERN]{European Organization for Nuclear Research (CERN), CH-1211 Geneve 23, Switzerland}
\address[USTC]{State Key Laboratory of Particle Detection and Electronics, University of Science and Technology of China, Hefei 230026, China}
\address[AUTH]{Department of Physics, Aristotle University of Thessaloniki, Thessaloniki, Greece}
\address[NSCR]{Institute of Nuclear and Particle Physics, NCSR Demokritos, 15341 Agia Paraskevi, Attiki, Greece}
\address[NTUA]{National Technical University of Athens, Athens, Greece}
\address[LIPP]{Laborat\'orio de Instrumenta\c{c}\~ao e F\'isica Experimental de Part\'iculas, Lisbon, Portugal}
\address[RD51]{RD51 collaboration, European Organization for Nuclear Research (CERN), CH-1211 Geneve 23, Switzerland}
\address[LIDYL]{LIDYL, CEA, CNRS, Université Paris-Saclay, F-91191 Gif-sur-Yvette, France}
\address[LIST]{CEA-LIST, Diamond Sensors Laboratory, CEA Saclay, F-91191 Gif-sur-Yvette, France}
\address[HIP]{Helsinki Institute of Physics, University of Helsinki, 00014 Helsinki, Finland}
\address[IGFAE]{Instituto Galego de F\'isica de Altas Enerx\'ias (IGFAE), Universidade de Santiago de Compostela, Spain}

\fntext[Veenhof]{Also at National Research Nuclear University MEPhI, Kashirskoe Highway 31, Moscow, Russia;
and Department of Physics, Uluda\u{g} University, 16059 Bursa, Turkey.}
\fntext[Virginia]{Also at University of Virginia}

\begin{abstract}
The PICOSEC detection concept consists in a ``two-stage'' Micromegas detector coupled to a Cherenkov radiator
and equipped with a photocathode.
A proof of concept has already been tested: a single-photoelectron response of 76\,ps has been measured
with a femtosecond UV laser at CEA/IRAMIS, while a time resolution of 24\,ps
with a mean yield of 10.4 photoelectrons has been measured for 150\,GeV muons at the CERN SPS H4 secondary line.
This work will present the main results of this prototype and the performance
of the different detector configurations tested in 2016-18 beam campaigns:
readouts (bulk, resistive, multipad) and photocathodes (metallic+CsI, pure metallic, diamond).
Finally, the prospects for building a demonstrator based on PICOSEC detection concept
for future experiments will be discussed.
In particular, the scaling strategies for a large area coverage with a multichannel readout plane,
the R\&D on solid converters for building a robust photocathode and the different resistive configurations for a robust readout.
\end{abstract}

\begin{keyword}
picosecond timing, MPGD, Micromegas, photocathodes, timing algorithms
\end{keyword}
\end{frontmatter}

Fast timing is a key feature for particle identification
through particle mass measurement at high-energy colliders and in nuclear physics facilities.
More specifically, tens of picoseconds in precision timing are needed for the High Luminosity upgrade of LHC (HL-LHC)
in order to mitigate the foreseen pile-up background of ATLAS/CMS experiments,
which could reach in some cases 150-200 vertices~\cite{White:2013taa}.
Various types of detectors are being developed to reach such precision~\cite{Vavra:2017jv}.
In solid-state field, Silicon photomultipliers (SiPMs)~\cite{Gundacker:2013ywa},
Low Gain Avalanche Detectors (LGAD)~\cite{Pellegrini:2014lki},
Micro-Channel Plate Photo-Multiplier (MCP-PMT)~\cite{Ronzhin:2015idh}
or Ultra Fast Silicon Detectors (UFSD)~\cite{Cartiglia:2015iua} are some examples.
In gaseous detectors, the PICOSEC detection concept, based on Micro-Pattern Gaseous Detectors (MPGD),
has recently reached a sub-25 picosecond timing precision
for Minimum Ionization Particles (MIP) of 150\,GeV muons~\cite{Bortfeldt:2017jb}.
This manuscript reviews the main results of the first PICOSEC prototype
and the scaling strategies to build a demonstrator for a future experiment or upgrade.

\section{Detection concept}
\label{sec:Detector}
\begin{figure}[htb!]
\centering
\includegraphics[width=0.99\linewidth]{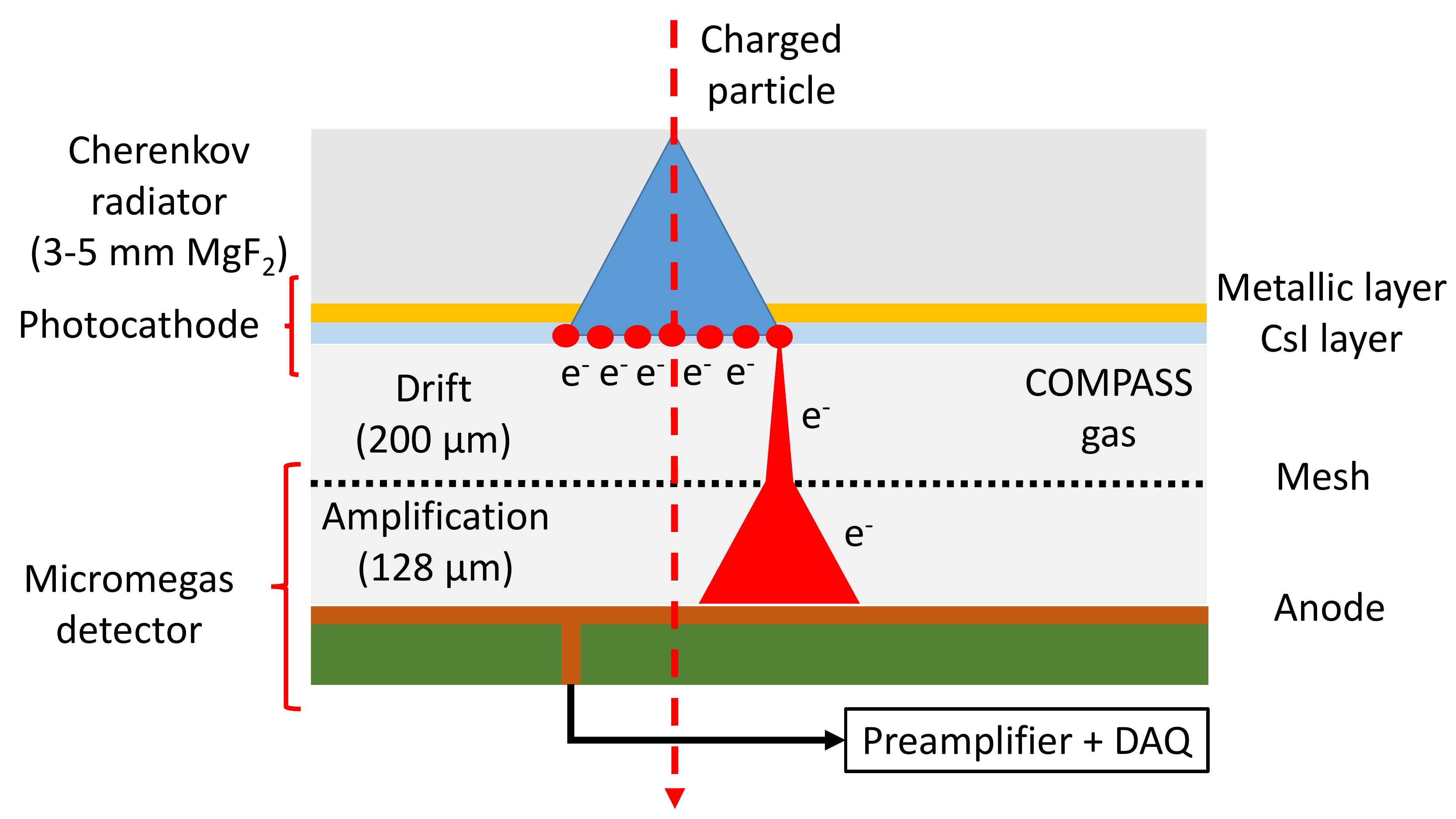}
\caption{Schema of the first PICOSEC prototype.
In beam tests, a charged particle produces UV photons when passing through the Cherenkov radiator.
These photons are then absorbed at the photocathode and partially converted to photoelectrons.
In laser tests, the laser impacts on the photocathode and produces single photoelectrons.
In both cases, photoelectrons are amplified in the two stages of the Micromegas detector (drift and amplification gaps),
and the secondary electrons induce a fast signal in the anode.
The Micromegas detector is filled with COMPASS gas at 1 bar absolute pressure.}
\label{fig:DetSchema}
\end{figure}

\begin{figure}[htb!]
\centering
\includegraphics[width=0.90\linewidth]{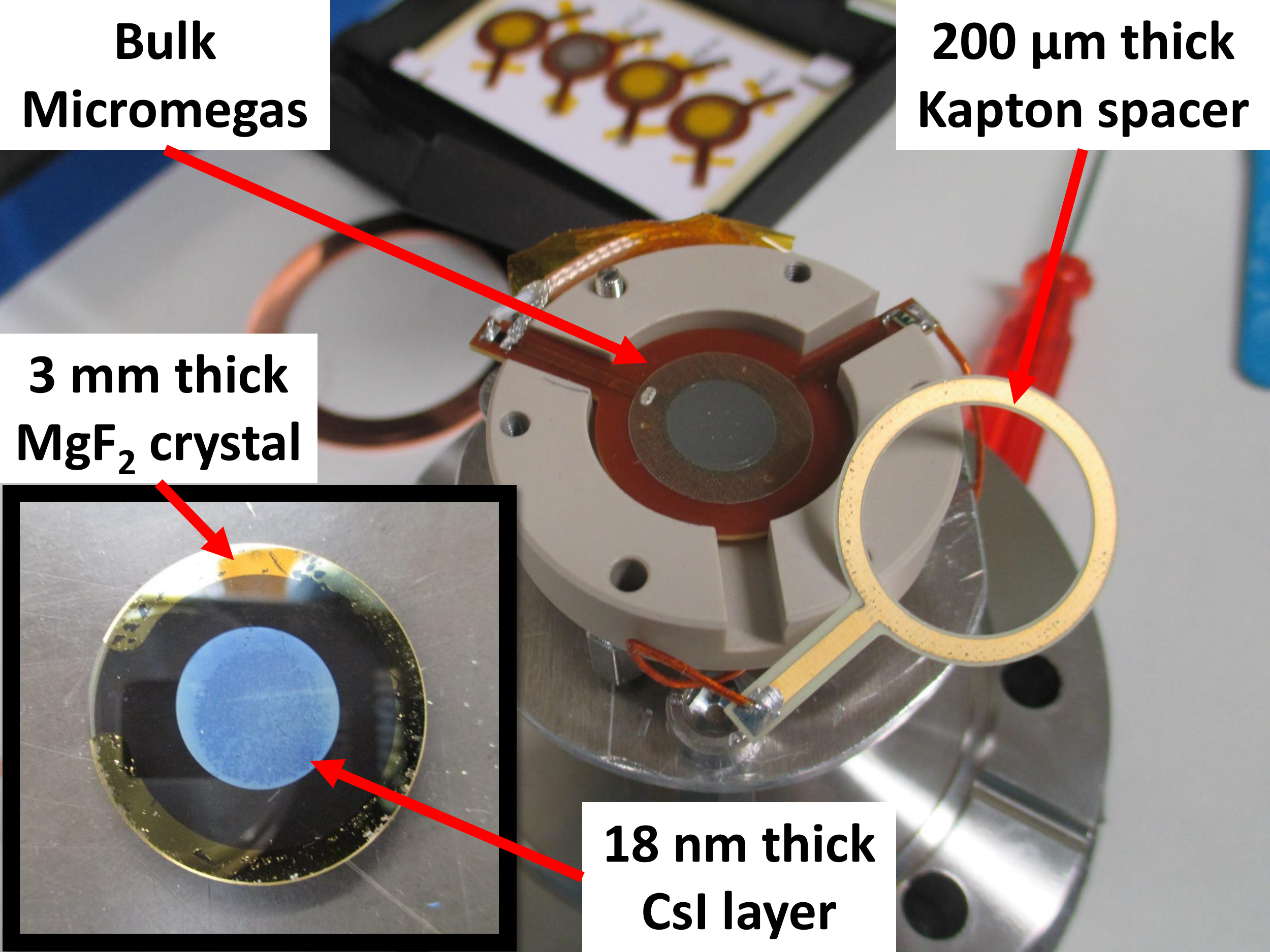}
\caption{Photograph of the first PICOSEC prototype. The bulk Micromegas detector has a 1\,cm diameter active area
and its 128\,\textmu m thick amplification gap is defined by six pillars, arranged in an hexagonal pattern.
The mesh and anode voltages are supplied by two strip-lines onto which two coaxial cables are soldered outside the chamber.
Inset: Photograph of the Cherenkov radiator and photocathode.
It is composed of a MgF$_2$ crystal (two-inches radius and 3\,mm thickness),
on top of which a 5.5\,\textmu m thick Chromium layer is deposited to polarize the crystal
and a CsI film (12\,mm-diameter and 18\,nm thickness) that works as photocathode.}
\label{fig:DetPhoto}
\end{figure}

The PICOSEC detector concept~\cite{Bortfeldt:2017jb} consists of a two-stage amplification Micromegas detector
coupled to a Cherenkov radiator coated with a photocathode, as shown in Fig.~\ref{fig:DetSchema}.
A charged particle crossing the radiator produces Cherenkov light in the extreme Ultra Violet (UV) wavelength (below 200\,nm),
which is simultaneously converted into electrons at the photocathode (a 18\,nm CsI layer in the first prototype).
The photocathode is deposited on a metallic layer, typically a 5.5\,nm-thick chromium layer,
which works as cathode of the first stage of the Micromegas (called \textit{drift gap} in literature).
The primary photo-electrons are preamplified in this stage, partially traverse the Micromegas mesh
and are finally amplified in the second stage (called \textit{amplification gap}).
In the first prototype (Fig.~\ref{fig:DetPhoto}), the drift gap is 200\,\textmu m thick and is defined by circular kapton spacers,
while the amplification gap is 128\,\textmu m thick and is defined by only 6 pillars.
The two gaps are filled with COMPASS gas (80\%Ne + 10\%C$_2$H$_6$ + 10\%CF$_4$) at 1\,bar absolute pressure.

\begin{figure}[htb!]
\centering
\includegraphics[width=0.99\linewidth]{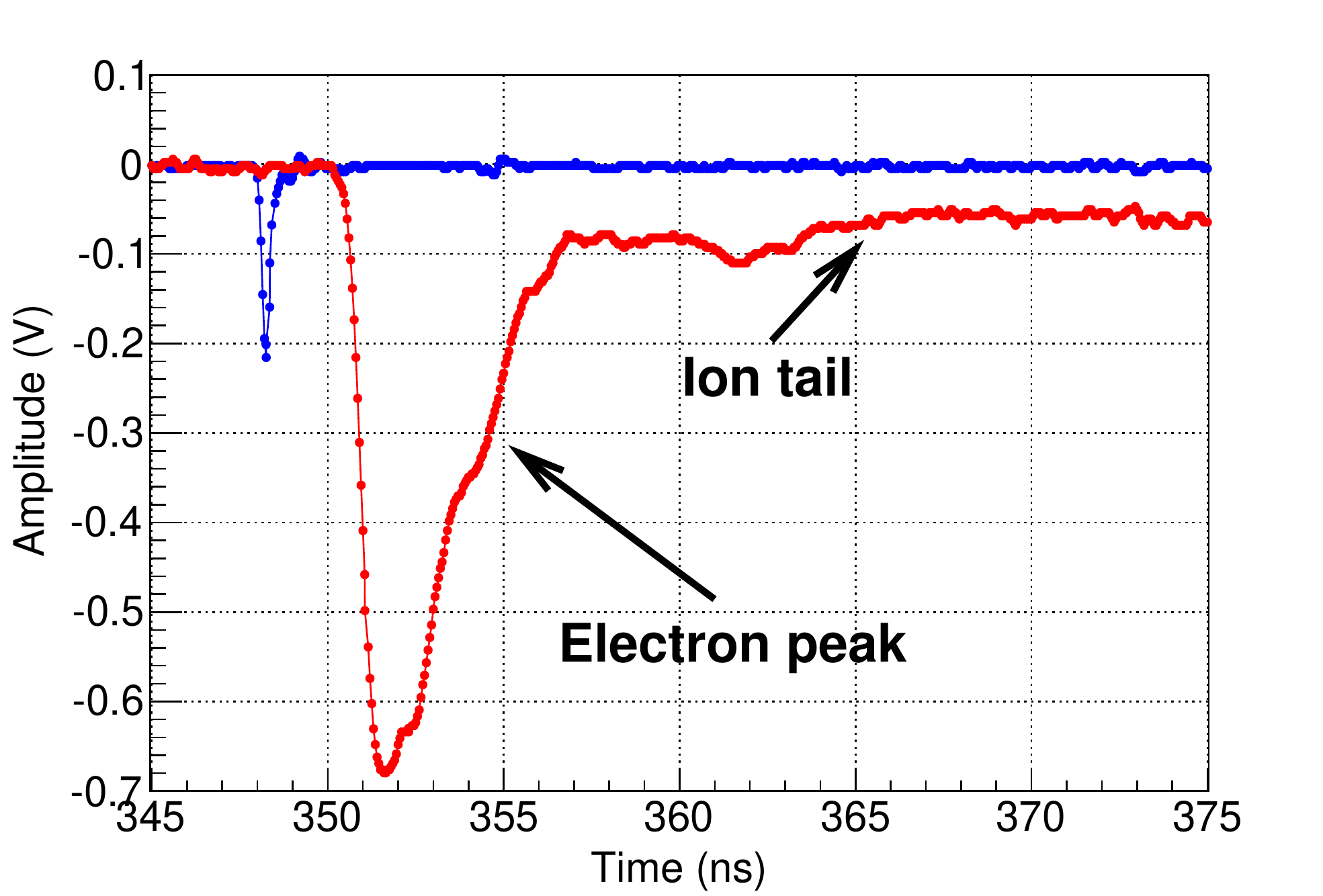}
\caption{An induced signal in the PICOSEC detector by a 150\,GeV muon (red points),
recorded together with the one produced in a Microchannel Plate PMT (blue points).
The PICOSEC signal contains a fast component produced by the fast movement of avalanche electrons to the Micromegas anode,
(called electron-peak)
and a slower component generated by ions drifting to the Micromegas mesh (called ion tail).}
\label{fig:PicosecSignal}
\end{figure}

The arrival of the amplified electrons at the Micromegas anode generates a fast signal
(with a risetime of $\sim$0.5\,ns) referred to as the electron-peak,
while the movement of the ions, produced in the amplification gap, to the mesh
generates a slower component - ion-tail ($\sim$100\,ns).
An example of the induced signal by a 150\,GeV muon is illustrated in Fig.~\ref{fig:PicosecSignal}.

\section{Timing results}
\label{sec:Results}
\begin{figure}[htb!]
\centering
\includegraphics[width=0.99\linewidth]{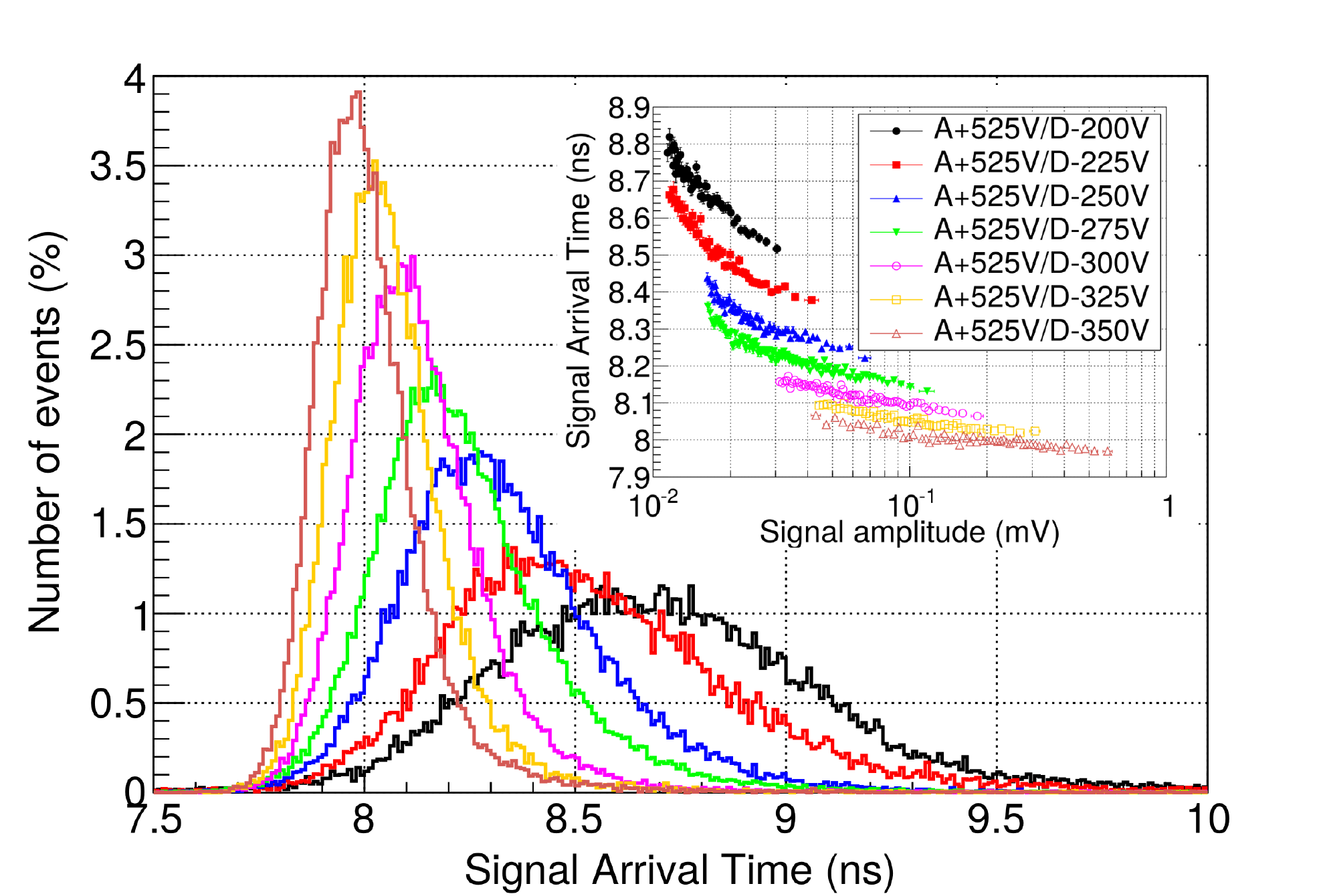}
\caption{An example of the Signal Arrival Time distributions generated by single photoelectrons in COMPASS gas
at 1\,bar absolute pressure for an anode voltage of 525\,V and drift voltages from 200 to 350\,V.
Inset: Mean of the signal arrival time as a function of the signal amplitude in the same conditions.}
\label{fig:SATLaser}
\end{figure}

The time response of the first PICOSEC prototype has been measured for single photoelectrons
with the help of a femtolaser of the CEA/IRAMIS laboratory.
In our setup, the laser beam is split into two equal parts,
one arrives directly at the prototype and the other at a fast photodiode, with a time resolution of $\sim$13 ps.
The intensity of the laser arriving at the detector is reduced by light attenuators,
so that the charge distribution is compatible with single photoelectron.
The PICOSEC detector signal is preamplified by a CIVIDEC module (2\,GHz, 40\,dB)
before being digitized and registered together with the photodiode
signal by a 2.5\,GHz oscilloscope at a rate of 20\,GSamples/s (i.e. one sample every 50\,ps).
The temporal distance between the two signals (called \textit{Signal Arrival Time} or SAT) is calculated by a Constant Fraction method.
An example of the resulting SAT distributions is shown in Fig.~\ref{fig:SATLaser} for a fixed anode voltage.
This figure illustrates: 1) the improvement in timing with the drift field;
and 2) the correlation between the SAT and the signal amplitude, which causes a tail at high SAT values.
These observations do not have an electronic origin, as the signal shape is the same for different signal sizes,
but can be explained by the charge amplification in the first Micromegas detector stage,
as shown in a detailed detector response simulation~\cite{Paraschou:2017kp}.
The detector time resolution at each operation point has been derived from the SAT-amplitude correlations,
leading to a best value of 76.0 $\pm$ 0.4\,ps~\cite{Bortfeldt:2017jb}.

\begin{figure}[htb!]
\centering
\includegraphics[width=0.99\linewidth]{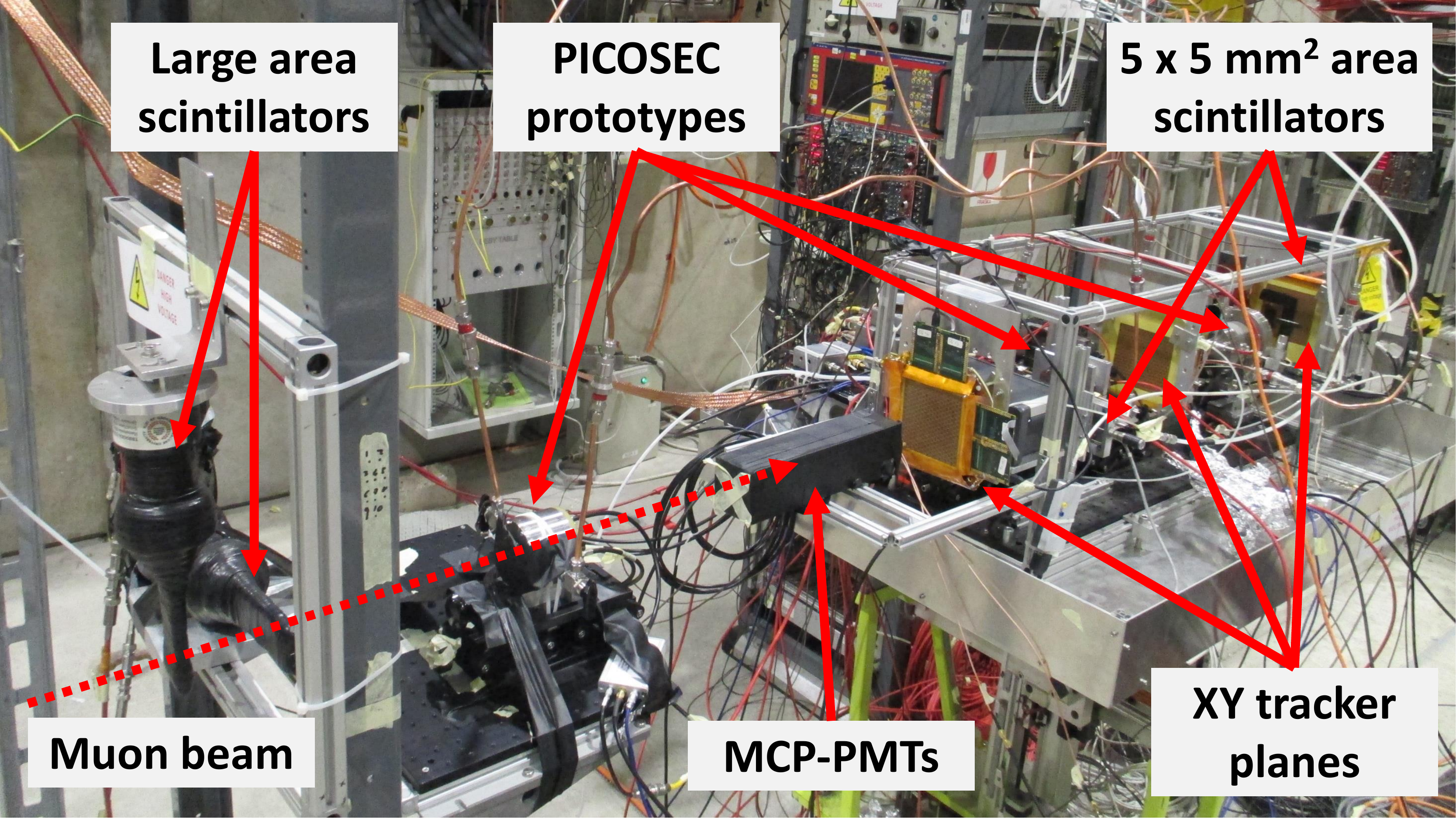}
\caption{Photograph of the beam telescope during the beam test in May 2018.
The incoming beam enters from the left side of the picture. Events are triggered either
by the coincidence of two $5 \times 5$\,mm$^2$ area scintillators in anti-coincidence with a veto scintillator at the end of the telescope
or by two large area scintillators in coincidence situated at the front part of the telescope.
Three XY tracker planes provide tracking information of the incoming charge particles,
while two MCPs give the time reference for the PICOSEC prototypes.}
\label{fig:BeamSetupPhoto}
\end{figure}

\begin{figure}[htb!]
\centering
\includegraphics[width=0.90\linewidth]{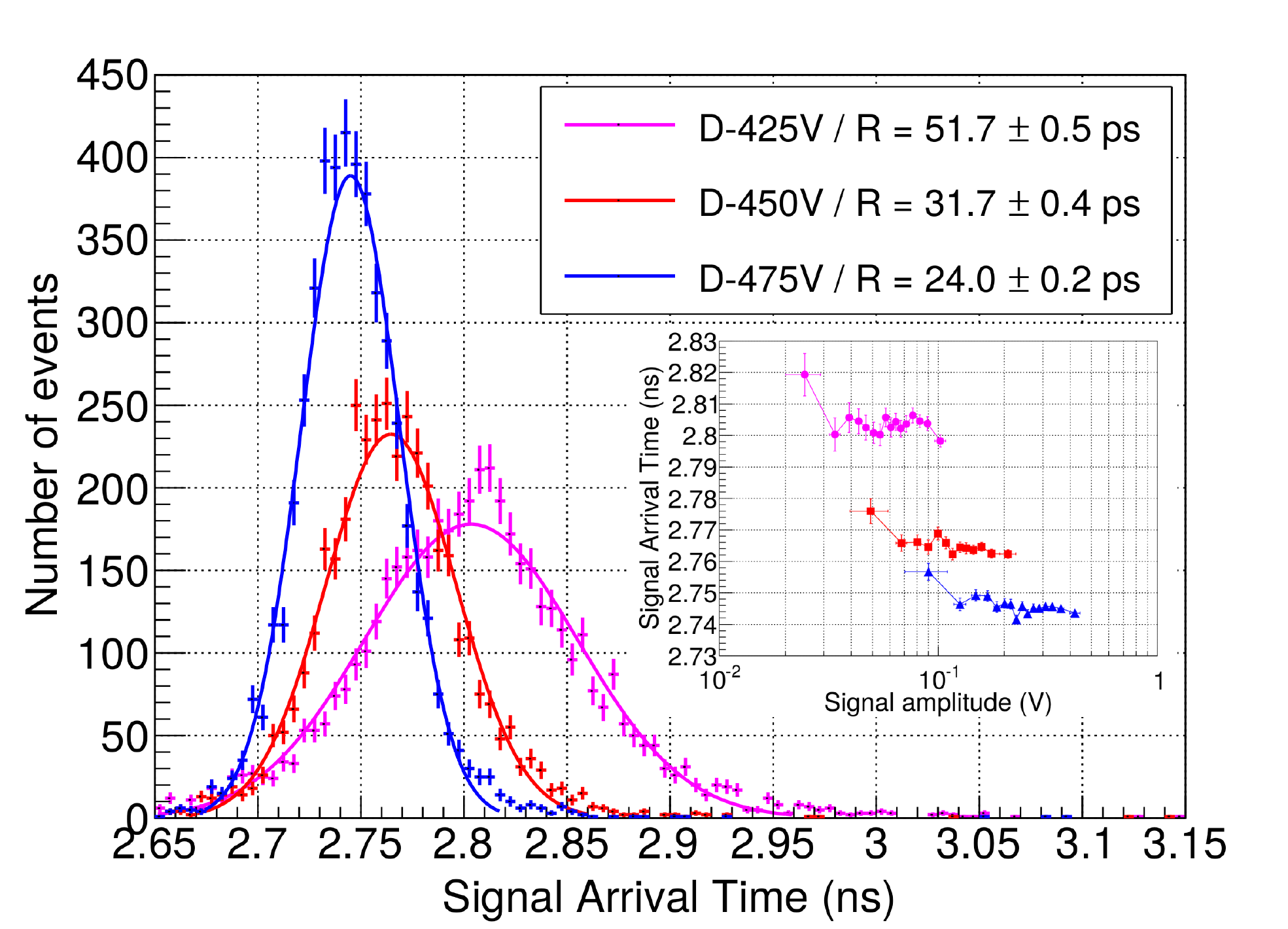}
\caption{An example of the Signal Arrival Time distributions generated by 150\,GeV muons in COMPASS gas
and the superimposed fit with a Gaussian function
for an anode voltage of 275\,V and drift voltages from 425 to 475\,V.
Inset: Mean of the signal arrival time as a function of the signal amplitude.}
\label{fig:SATBeam}
\end{figure}

The time response of the PICOSEC detector to 150\,GeV muons has been measured
in several beam tests in 2016-18 at the CERN SPS H4 secondary beamline.
The beam telescope (Fig.~\ref{fig:BeamSetupPhoto}) consists of a tracking system
composed of three triple-Gas Electron Multipliers (GEMs) 
and with a spatial resolution better than 40\,\textmu m;
one MCP-PMT~\cite{Ronzhin:2015idh} as time reference with a time resolution better than 5\,ps~\cite{Sohl:2018ls};
different scintillators to select tracks and to veto showers; and up to five prototypes.
As in the laser setup, the PICOSEC signal is preamplified and registered together with the MCP-PMT signal,
as well as one digitized event number from the tracking system.
An example of the SAT distributions of the first PICOSEC prototype is shown in Fig.~\ref{fig:SATBeam}.
This figures illustrates: 
1) the improvement in timing with the drift field;
2) the absence of correlation between the SAT and the amplitude, due to high drift voltages;
and 3) a best value for time resolution of 24.0 $\pm$ 0.2\,ps, as reported in~\cite{Bortfeldt:2017jb}.
The mean number of photoelectrons per muon was also calculated comparing
the electron-peak charge distribution by the one generated by single photoelectrons (measured by a UV-lamp calibration),
giving a value of 10.4 $\pm$ 0.4.

\section{On going R\&D for a demonstrator}
\label{sec:Demonstrator}
\begin{figure}[htb!]
\centering
\includegraphics[width=0.85\linewidth]{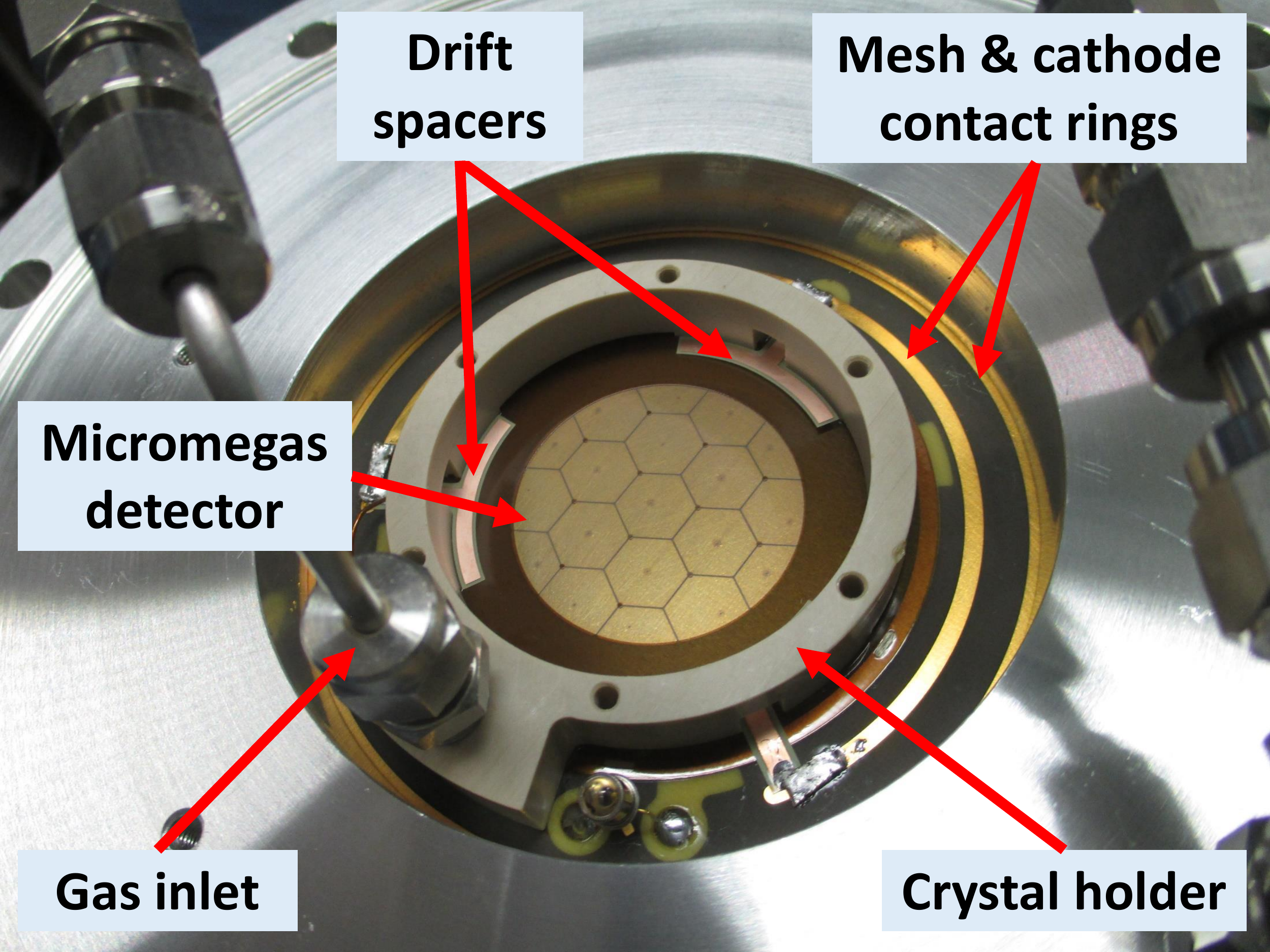}
\caption{Picture of the Multipad PICOSEC detector mounted on a flange of the chamber,
where several components are marked: the Micromegas detector (with its 19 hexagonal pads), the drift spacers,
the two contact rings for the mesh and cathode voltages and the PEEK structure that holds a 2 inches-diameter crystal
with a photocathode.}
\label{fig:MultipadDetector}
\end{figure}

\begin{figure}[htb!]
\centering
\includegraphics[width=0.90\linewidth]{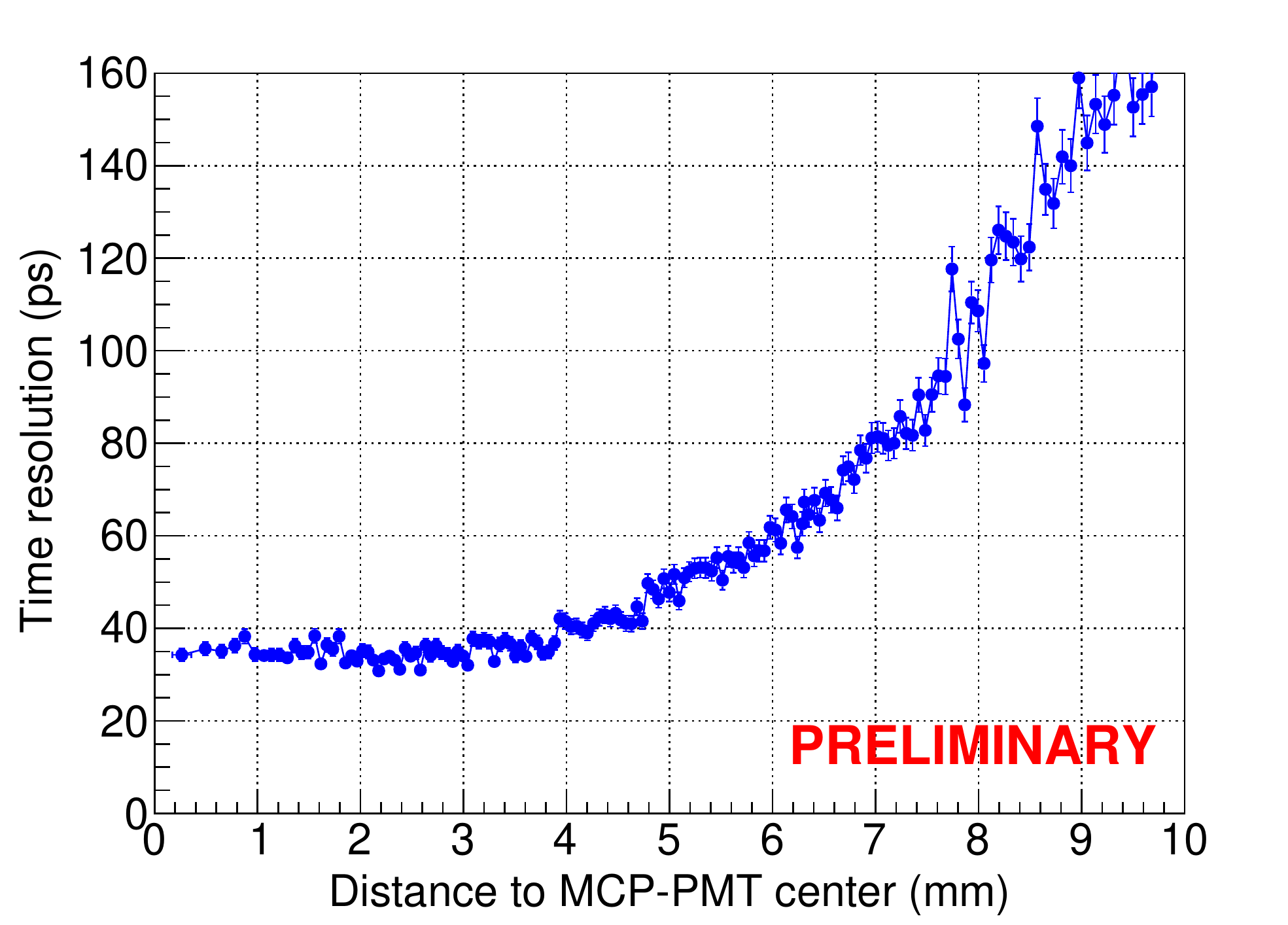}
\caption{Dependence of the time resolution of the Multipad detector 
on the distance of the impact point to MCP-PMT center
after combining the signals of the three central anode pads.}
\label{fig:TDistTResRadius}
\end{figure}

Several components of the first PICOSEC prototype must gain in robustness and scalability
before a demonstrator could be proposed to an experiment. This is the aim of several on going R\&Ds,
briefly discussed in this section:
\begin{enumerate}
 \item Resistive Micromegas detectors~\cite{Wotschack:2013ola} do not show any degradation of the signal efficiency
 and the spatial resolution in pion beams with respect to non resistive detectors~\cite{Manjarres:2012mp}.
 Two configurations have been already tested: the resistive strip one~\cite{Alexopoulos:2011zz}
 and the floating strip one~\cite{Bortfeldt:2014vvt}.
 In both cases, the best value for time resolutions is 28-40\,ps, slightly worse than the reference value,
 but with a proven operation and stability in 150\,GeV pion beam over hours.
 
 \item The Multipad detector (Fig.~\ref{fig:MultipadDetector}) is the first prototype
 to test the scalability of PICOSEC detection concept.
 It has a 36\,mm diameter active area divided in 19 hexagonal anode pads.
 The 19 anode pads, the mesh and cathode electrodes
 are routed to the back side of the Printed Circuit Board (PCB), where they are connected to the electronics.
 A woven stainless-steel mesh is laminated on the PCB to make a bulk detector (128\,\textmu m thick amplification gap).
 The 200\,\textmu m thick drift gap is defined by ring spacers and the 2 inches diameter crystal is held by a PEEK structure.
 
 The detector was tested in October 2017 beam tests in two configurations:
 in the first one, the MCP-PMT, the central pad and a 5 $\times$ 5 mm$^2$ scintillator were aligned
 to study the timing of a single pad for different voltage configurations; in the second one, the MCP-PMT center was situated
 at the intersection of three pads, the pads were covered by a 4 $\times$ 5 cm$^2$ scintillator area
 and $\sim 10^6$ events were acquired at the optimum operation point in order to combine the timing of the three pads.
 In both cases, preliminary results show a time resolution of $\sim$36\,ps.
 The result of the second case is limited by the MCP optimum performance area (Fig.~\ref{fig:TDistTResRadius}),
 which extends up to a radius of 5.5\,mm~\cite{Sohl:2018ls}.
 
 \item CsI shows a high quantum efficiency as a photocathode
 (of more than 25\% at 180\,nm wavelength~\cite{LU1994135})
 but is hydrophobic and less resistant to ion-backflow~\cite{Breskin:1995kr}.
 Different alternatives are being explored:
 pure metallic photocathodes, diamond-based photocathodes or protection layers for the CsI.
 Preliminary results have shown a time resolution of $\sim$60\,ps for the first two types (5\,mm thick MgF$_2$ + 10\,nm Al
 and 3\,mm thick MgF$_2$ + 20\,nm Diamond Like Carbon), and $\sim$2 photoelectrons per muon.
 \item The commercial preamplifier will be replaced by on-board electronics with spark protection in a new prototype;
 while the substitution of oscilloscope as DAQ with ultra-fast digitizers like SAMPIC~\cite{Delagnes:2015oda} is also being studied.
\end{enumerate}

\section{Conclusions}
\label{sec:Conclusions}
The PICOSEC detection concept, composed of a two-stage Micromegas detector coupled to a Cherenkov radiator
and equipped with a photocathode, has been presented.
The first prototype has shown a timing resolution of about 76\,ps
for single photoelectrons and about 24\,ps for 150\,GeV muons.
To improve the robustness and scalability of different components, a R\&D program has been launched.
First results with resistive Micromegas, in multipad configuration and with different photocathodes are promising.

\section*{Acknowledgments}
We acknowledge the financial support of the RD51 collaboration, in the framework of RD51 common projects,
the Cross-Disciplinary Program on Instrumentation and Detection of CEA, the French Alternative Energies and Atomic Energy Commission;
and the Fundamental Research Funds for the Central Universities of China.
We also thank K. Kordas for valuable suggestions concerning the analysis of the data.
J.~Bortfeldt acknowledges the support from the COFUND-FP-CERN-2014 program (grant number 665779).
M.~Gallinaro acknowledges the support from the Funda\c{c}\~ao para a Ci\^{e}ncia e a Tecnologia (FCT), Portugal
(grants IF/00410/2012 and CERN/FIS-PAR/0006/2017).
D.~Gonz\'alez-D\'iaz acknowledges the support from MINECO (Spain) under the Ramon y Cajal program (contract RYC-2015-18820).
F.J.~Iguaz acknowledges the support from the Enhanced Eurotalents program (PCOFUND-GA-2013-600382).
S. White acknowledges partial support through the US CMS program under DOE contract No. DE-AC02-07CH11359.

\bibliographystyle{JHEP}
\bibliography{20180802_FJIguaz_PICOSEC}

\providecommand{\href}[2]{#2}\begingroup\raggedright\begin{thebibliography}{10}

\bibitem{White:2013taa}
S.~White, \emph{{``Experimental Challenges of the European Strategy for
  Particle Physics''}},  in \emph{{Proceedings, International Conference on
  Calorimetry for the High Energy Frontier (CHEF 2013): Paris, France, April
  22-25, 2013}}, pp.~118--127, 2013.
\newblock \href{https://arxiv.org/abs/1309.7985}{{\ttfamily 1309.7985}}.

\bibitem{Vavra:2017jv}
J.~Vavra, \emph{{PID techniques: Alternatives to RICH methods}},
  \href{http://dx.doi.org/10.1016/j.nima.2017.02.075}{\emph{Nucl. Instrum.
  Meth.} {\bfseries A876} (2017) 185--193}.

\bibitem{Gundacker:2013ywa}
S.~Gundacker et~al., \emph{{SiPM time resolution: From single photon to
  saturation}}, \href{http://dx.doi.org/10.1016/j.nima.2013.01.047}{\emph{Nucl.
  Instrum. Meth.} {\bfseries A718} (2013) 569--572}.

\bibitem{Pellegrini:2014lki}
G.~Pellegrini et~al., \emph{{Technology developments and first measurements of
  Low Gain Avalanche Detectors (LGAD) for high energy physics applications}},
  \href{http://dx.doi.org/10.1016/j.nima.2014.06.008}{\emph{Nucl. Instrum.
  Meth.} {\bfseries A765} (2014) 12--16}.

\bibitem{Ronzhin:2015idh}
A.~Ronzhin, S.~Los, E.~Ramberg, A.~Apresyan, S.~Xie, M.~Spiropulu et~al.,
  \emph{{Study of the timing performance of micro-channel plate photomultiplier
  for use as an active layer in a shower maximum detector}},
  \href{http://dx.doi.org/10.1016/j.nima.2015.06.006}{\emph{Nucl. Instrum.
  Meth.} {\bfseries A795} (2015) 288--292}.

\bibitem{Cartiglia:2015iua}
N.~Cartiglia et~al., \emph{{Design optimization of ultra-fast silicon
  detectors}}, \href{http://dx.doi.org/10.1016/j.nima.2015.04.025}{\emph{Nucl.
  Instrum. Meth.} {\bfseries A796} (2015) 141--148}.

\bibitem{Bortfeldt:2017jb}
J.~Bortfeldt et~al., \emph{{PICOSEC: Charged particle timing at sub-25
  picosecond precision with a Micromegas based detector}},
  \href{http://dx.doi.org/10.1016/j.nima.2018.04.033}{\emph{accepted in Nucl.
  Instrum. Meth. A} (2018) }.

\bibitem{Paraschou:2017kp}
K.~Paraschou and S.~Tzamarias, ``A data driven simulation study of the timing
  effects observed with the picosec micromegas detector.''
  \url{https://indico.cern.ch/event/676702/contributions/2809871/attachments/1574857/2486512/Konstantinos_RD51_miniweek.pdf},
  Dec. 13, 2017.

\bibitem{Sohl:2018ls}
L.~Sohl, \emph{{Spatial time resolution of MCP–PMTs as a t$_0$-reference}},
  {\emph{submitted to Nucl. Instrum. Meth. A} (2018) }.

\bibitem{Wotschack:2013ola}
J.~Wotschack, \emph{{The development of large-area Micromegas detectors for the
  ATLAS upgrade}},
  \href{http://dx.doi.org/10.1142/S0217732313400208}{\emph{Mod. Phys. Lett.}
  {\bfseries A28} (2013) 1340020}.

\bibitem{Manjarres:2012mp}
{\scshape MAMMA} collaboration, J.~Manjarres et~al., \emph{{Performances of
  Anode-resistive Micromegas for HL-LHC}},
  \href{http://dx.doi.org/10.1088/1748-0221/7/03/C03040}{\emph{JINST}
  {\bfseries 7} (2012) C03040}.

\bibitem{Alexopoulos:2011zz}
T.~Alexopoulos et~al., \emph{{A spark-resistant bulk-micromegas chamber for
  high-rate applications}},
  \href{http://dx.doi.org/10.1016/j.nima.2011.03.025}{\emph{Nucl. Instrum.
  Meth.} {\bfseries A640} (2011) 110--118}.

\bibitem{Bortfeldt:2014vvt}
J.~Bortfeldt, \emph{The Floating Strip Micromegas Detector}.
\newblock Springer, 2015,
  \href{http://dx.doi.org/10.1007/978-3-319-18893-5}{10.1007/978-3-319-18893-5}.

\bibitem{LU1994135}
C.~Lu and K.~McDonald, \emph{Properties of reflective and semitransparent csi
  photocathodes},
  \href{http://dx.doi.org/10.1016/0168-9002(94)90543-6}{\emph{Nucl. Instrum.
  Meth.} {\bfseries A343} (1994) 135--151}.

\bibitem{Breskin:1995kr}
A.~Breskin, \emph{{CsI UV photocathodes: History and mystery}},
  \href{http://dx.doi.org/10.1016/0168-9002(95)01145-5}{\emph{Nucl. Instrum.
  Meth.} {\bfseries A371} (1996) 116--136}.

\bibitem{Delagnes:2015oda}
E.~Delagnes, D.~Breton, H.~Grabas, J.~Maalmi and P.~Rusquart, \emph{{Reaching a
  few picosecond timing precision with the 16-channel digitizer and timestamper
  SAMPIC ASIC}},
  \href{http://dx.doi.org/10.1016/j.nima.2014.12.042}{\emph{Nucl. Instrum.
  Meth.} {\bfseries A787} (2015) 245--249}.

\end{thebibliography}\endgroup
\end{document}